\documentclass[aps,prd,twocolumn,showpacs,groupedaddress,eqsecnum,floatfix]{revtex4-1}
\usepackage{graphicx,amsmath,amssymb,epsfig}
\begin{document}
\title{AdS/CFT Energy Loss in \\ Time-Dependent String Configurations}
\author{Andrej Ficnar}
\email{aficnar@phys.columbia.edu}
\affiliation{Department of Physics, Columbia University, New York, NY 10027, USA}
\date{\today}

\begin{abstract}
We analyze spacetime momentum currents on a classical string worldsheet, study their generic connection via AdS/CFT correspondence to the instantaneous energy loss of the dual field theory degrees of freedom and suggest a general formula for computing energy loss in a time-dependent string configuration. Applying this formula to the case of falling strings, generally dual to light quarks, reveals that the energy loss does not display a well-pronounced Bragg peak at late times, as previously believed. Finally, we comment on the possible implications of this result to the jet quenching phenomena in heavy ion collisions.
\end{abstract}
\maketitle


\section{Introduction}

Gauge/gravity duality has been a very insightful tool used to study many properties of strongly coupled non-Abelian plasmas, especially after many indications that the quark-gluon plasma created in heavy ion collisions at the RHIC collider is a strongly coupled system \cite{miklos-larry}. The AdS/CFT correspondence \cite{maldacena,witten,gkb} is a duality between a (3+1)-dimensional $\mathcal{N}=4$ $SU(N_c)$ super-Yang-Mills (SYM) gauge theory and type IIB string theory on $AdS_5\times S^5$ spacetime. Using this conjecture and by taking the limit $N_c\gg \lambda \gg 1$, one can study this gauge theory at strong coupling by studying classical, two-derivative (super)gravity. Due to phenomenological differences between thermal $\mathcal{N}=4$ SYM plasma and finite-temperature QCD, it is important to consider gravity duals to nonconformal gauge theories \cite{kiritsis-prl,kiritsis-comparison,gubser-nellore,gubser-prl,jorge-polyloops,proceeding1,proceeding2,our-future-paper}. However, since the information about the medium (on the field theory side) is fully encoded in the spacetime metric (on the string side), in this work we will keep our results as general as possible by considering a general metric $G_{\mu\nu}$ and only use the $AdS_5$ metric for final numerical evaluations.

In recent years, an important application of the AdS/CFT correspondence has been to study the phenomenon of jet quenching \cite{jq1,jq2,jq3,jq4,jq5} in strongly coupled systems, especially after the pioneering work of \cite{drag-force} and \cite{herzog}, who studied energy loss of heavy quarks in a strongly coupled $\mathcal{N}=4$ SYM plasma. To study the plasma at a finite temperature $T$, one introduces a black hole in the $AdS_5$ geometry with an event horizon at some radial coordinate $r_h$, proportional to $1/T$ \cite{witten}. Then, one introduces degrees of freedom in the fundamental representation (``quarks'') in SYM by introducing a D7-brane in the $AdS$-BH geometry, which spans from $r=0$ (boundary) to some $r=r_m$ \cite{karch-katz}; on the field theory side, this procedure corresponds to the introduction of an $\mathcal{N}=2$ hypermultiplet whose mass $m_Q$ is proportional to $1/r_m$. Dressed quarks are then dual to strings in the bulk with one or both endpoints on the D7-brane and the physics of the quark energy loss (on the field theory side) will be directly related to the dynamics of these strings. In the large $N_c$ and large $\lambda$ limit, one can neglect the backreaction of the metric due to the introduction of the string (the probe approximation) and neglect the quantum corrections to the motion of the strings. This means that we need to study the motion of classical strings in the background given by a spacetime metric $G_{\mu\nu}$. For a review of AdS/CFT correspondence and especially its applications to heavy ion phenomenology, the reader is referred to \cite{review1,review2,review3}.

Recently, in the light of RHIC results, as well as the new LHC results \cite{lhc1,lhc2} on the suppression of light hadrons in AA collisions, a more consistent grasp on the energy loss of light quarks in gauge/gravity duality has become necessary, in order to be able to compute jet energy loss observables such as the nuclear modification factor $R_{AA}$ and the elliptic flow parameter $v_2$ and directly compare them to the experimental results \cite{miklos-abc}. If we wish to describe a strongly coupled medium with light quarks, the D7-brane will fill the entire $AdS$-BH geometry. On this D7-brane we can then study open strings whose endpoints source a D7 gauge field which in turn ``induces'' (in the sense of the field/operator correspondence) an image baryon density current in the field theory. In other words, these open strings will represent dressed $q\bar{q}$ pairs on the field theory side. Then the main idea, advocated in \cite{chesler}, is that by studying the free motion of these falling strings, we can study the energy loss of light quarks. 

This application is just one example of the need to understand the details of energy loss in (explicitly) time-dependent string configurations, such as the falling strings. The hope is that, by examining such configurations, we can model phenomena associated with the energy loss more realistically, since in the quark gluon plasma  formed in heavy ion collisions, quarks slow down and phenomena such as the instantaneous energy loss are expected to depend on the details of this nonstationary motion. Heavy quarks in such nonstationary situations have already been studied in \cite{guijosa1,guijosa2}.

The authors of \cite{chesler} have elegantly obtained one of the first results for light quark energy loss in gauge/gravity duality. They have shown, by analyzing null geodesics in the $AdS$-BH spacetime and relating them to the energy of the falling string, that the maximum stopping distance of light quarks scales with energy as $\Delta x_{max}\sim E^{1/3}$. A similar result was obtained in \cite{gubser-gluon} for energy loss of adjoint degrees of freedom (``gluons'') in $\mathcal{N}=4$ SYM plasma. However, we emphasize that this maximum stopping distance is not a {\it typical} stopping distance of light quarks. It is, as such, a rather crude quantity, that can be used as a phenomenological guideline, but cannot be used to extract the instantaneous energy loss that enters in, for example, calculations of $R_{AA}$ or $v_2$ observables. To study the instantaneous energy loss, one needs to analyze the spacetime momentum currents on the string worldsheet, which, as demonstrated in \cite{chesler}, in case of falling strings become nontrivial, time-dependent quantities. In this paper we will extend the analysis of the worldsheet currents from \cite{chesler} and suggest a perhaps more appropriate definition of the energy loss (in the sense of its identification with particular current components), in which the details of the geometry on the worldsheet become important. This small, but crucial detail will lead to potentially important phenomenological consequences: as we will see, in particular, the instantaneous energy loss will not exhibit a well-pronounced late-time Bragg-like peak, as previously believed.


\section{Dynamics of classical strings}

Let us start by considering a classical string propagating in a five-dimensional spacetime described by the metric $G_{\mu\nu}$. The dynamics of the string is described by the Polyakov action:
\begin{equation}\label{st1}
S_P=-\frac{1}{4\pi\alpha'}\int d^2\sigma \sqrt{-h}h^{ab}(\partial_a X^\mu)(\partial_b X^\nu)G_{\mu\nu}\, ,
\end{equation}
where $\alpha'=l_s^2$, the squared fundamental string length; $\sigma^a=(\sigma,\tau)$ are the coordinates on the string worldsheet; $X^\mu(\sigma,\tau)$ are the spacetime coordinates of the string (the embedding functions); and $h_{ab}$ is the worldsheet metric, which is considered as a dynamical variable in this action (here $h\equiv \det(h_{ab})$). If we vary this action with respect to $h^{ab}$, we get
\begin{equation}\label{st2}
\gamma_{ab}=\frac{1}{2}h_{ab}(h^{cd}\gamma_{cd})\, ,
\end{equation}
where $\gamma_{ab}=G_{\mu\nu}\partial_a X^\mu \partial_b X^\nu$ is the induced worldsheet metric. Plugging this equation of motion in the Polyakov action \eqref{st1}, we obtain the usual Nambu-Goto string action, which means that these two actions are classically equivalent. 

The Polyakov action can thus be viewed as a classical field theory action of five scalar fields $X^\mu$ on a curved two-dimensional manifold described by the metric $h_{ab}$. This is the reason why we will be using that action instead of the much more common Nambu-Goto action, since in the derivation of the energy loss formula it will be necessary to consider coordinate transformations on the worldsheet and see how the worldsheet vectors and tensors change under them. In practice, this action will also be useful for numerical evaluations, since a clever choice of the worldsheet metric will greatly improve the stability of the numerics \cite{chesler}, as we will see later.

The equations of motion for the $X^\mu$ fields from the Polyakov action are
\begin{equation}\label{st3}
\partial_a\left[\sqrt{-h}h^{ab}G_{\mu\nu}\partial_bX^\nu\right]=\frac{1}{2}\sqrt{-h}h^{ab}(\partial_\mu G_{\nu\rho})\partial_aX^\nu\partial_bX^\rho\, .
\end{equation}
The expression in the brackets on the LHS are just the canonical momentum densities:
\begin{equation}\label{st4}
\Pi_\mu^a\equiv \frac{1}{\sqrt{-h}}\frac{\delta S_P}{\delta(\partial_a X^\mu)}=-\frac{1}{2\pi\alpha'}h^{ab}(\partial_b X^\nu)G_{\mu\nu}\, .
\end{equation}
This definition can differ by a factor of $\sqrt{-h}$ from the usual definition of these momenta found in the literature, which is here just to ensure that the quantity in \eqref{st4} is a proper worldsheet vector. From now on, we will assume that the spacetime metric is diagonal and we will consider only $\mu$'s such that the metric does not depend on $X^\mu$. In that case, equations of motion \eqref{st3} are in fact just the covariant conservation law for the momentum densities:
\begin{equation}\label{st5}
\partial_a\left[\sqrt{-h}\,\Pi^a_\mu\right]=\sqrt{-h}\nabla_a\Pi_\mu^a=0\, .
\end{equation}
In this case, the momentum densities are just the conserved Noether currents on the worldsheet, associated with the invariance of the action to the constant spacetime translations $X^\mu\to X^\mu + \epsilon^\mu$. Due to this origin, these worldsheet currents describe the flow of the $\mu$ component of the spacetime momentum of the string along the $a$ direction on the worldsheet \cite{martinec,zwiebach} and that is the reason why they are important in the study of energy loss in the field theory dual \cite{stress}. Because of that it is also necessary to understand how these currents transform under a generic change of coordinates on the worldsheet.

A general coordinate transformation on the worldsheet $(\sigma,\tau)\to(\tilde{\sigma},\tilde{\tau})$ can be defined by the following matrix:
\begin{equation}\label{st6}
\tilde{M}^a{}_b\equiv \frac{\partial\tilde{\sigma}^a}{\partial\sigma^b},\;\;\; M^a{}_b\equiv \frac{\partial\sigma^a}{\partial\tilde{\sigma}^b}=(\tilde{M}^{-1})^a{}_b\, .
\end{equation}
The worldsheet currents \eqref{st4} and the worldsheet metric then transform as proper worldsheet tensors:
\begin{eqnarray}\label{st7}
\tilde{h}_{ab}&=&M^c{}_a M^d{}_b h_{cd}\, ,\\\label{st8}
\tilde{\Pi}^a_\mu&=&\tilde{M}^a{}_b \Pi^b_\mu\, .
\end{eqnarray}
In practice, one solves the equations of motion \eqref{st3} with constraints \eqref{st2} and an appropriate set of boundary conditions in some parametrization where the numerics are well behaved (by choosing a convenient $h_{ab}$ at the beginning of the calculation) and then, using formulas \eqref{st7} and \eqref{st8}, transforms to some more ``physical'' coordinate system on the worldsheet (for example $\sigma=r$, the radial $AdS$ coordinate). From now on, we assume that, through this procedure, a specific parametrization $(\sigma, t)$ has been chosen, where $t$ is the physical time (i.e. we are in the static gauge $\tau=t$) and that $\sigma\in[0,\pi]$ for all $t$, i.e. the $\sigma$ coordinate parametrizes the string at some fixed time $t$.


\section{Worldsheet currents}

In general, whenever we have a covariant conservation law on a differentiable manifold, such as \eqref{st5}, one defines the charge (whose flow is described by $\Pi_\mu^a$) that passes through some hypersurface $\gamma$ as (following conventions in \cite{carroll})
\begin{equation}\label{wc1}
p_\mu^\gamma=-\int\limits_\gamma \star \Pi_\mu=-\int\limits_\gamma d\epsilon \, n_a \Pi_\mu^a\, ,
\end{equation}
where the second equation is written in terms of the one-form $\Pi_{\mu a}=h_{ab}\Pi_\mu^b$ and in the second equation, $d\epsilon$ is the induced volume element on the hypersurface $\gamma$ and $n_a$ is the unit vector field normal to the hypersurface. In our case, we have a two-dimensional manifold, and $\gamma$ represents an (open) curve on the worldsheet and $p_\mu^\gamma$ is the $\mu$ component of the spacetime momentum that flows through this curve. Here one should think of $\mu$ as merely an index that denotes different kinds of conserved currents on the worldsheet. Therefore, in our case, we can simply write
\begin{equation}\label{wc2}
p_\mu^\gamma=-\int\limits_\gamma ds\, n^b\Pi_\mu^a h_{ab}\, ,
\end{equation}
where $ds$ is the line element of the curve $\gamma$. \\

In our choice of coordinates on the worldsheet $(\sigma,t)$, we can now ask what is the total $\mu$ component of the momentum of the string at some fixed time $t$. This means that we need to take the curve $\gamma$ to be the curve of constant $t$, which means that its tangent vector $t^a$ is in the $\sigma$-direction, i.e. $t^a=(1,0)$, where the first entry is the $\sigma$ coordinate and the second the $t$ coordinate. The normal vector $n^a=(n^\sigma,n^\tau)$ can then be found by requiring its orthogonality to the tangent vector and the usual normalization condition:
\begin{equation}\label{wc3}
t\cdot n=0,\,\, n\cdot n=-1\, ,
\end{equation}
where the dot products are taken with the worldsheet metric $h_{ab}$. Using these two equations, we can solve for the components of the normal vector:
\begin{equation}\label{wc4}
n^a=\left(-\frac{1}{\sqrt{-h}}\frac{h_{\sigma\tau}}{\sqrt{h_{\sigma\sigma}}},\frac{\sqrt{h_{\sigma\sigma}}}{\sqrt{-h}}\right)\, .
\end{equation}
Finally, since for this particular curve we have $d\tau=0$, we can express the line element $ds$ as
\begin{equation}\label{wc5}
ds^2_\gamma=h_{ab}d\sigma^a d\sigma^b=h_{\sigma\sigma}d\sigma^2\, .
\end{equation}
Using \eqref{wc4} and \eqref{wc5} in \eqref{wc2}, we have
\begin{equation}\label{wc6}
p_\mu (t)=\int d\sigma \, \sqrt{-h}\,\Pi_\mu^\tau(\sigma,t)\, .
\end{equation}
Note that this is true no matter what the parametrization of the string is. The only requirement here is that we are dealing with a constant-$t$ curve. Similarly, we can repeat the same procedure for a constant-$\sigma$ curve, integrating over some period of time:
\begin{equation}\label{wc7}
p_\mu (\sigma,\Delta t)=\int\limits_{\Delta t} dt \, \sqrt{-h}\,\Pi_\mu^\sigma(\sigma,t)\, ,
\end{equation}
which then gives the momentum that has flown down the string (i.e. in the direction of increasing $\sigma$) at position $\sigma$ during the time $\Delta t$. Both \eqref{wc6} and \eqref{wc7} are the well-known formulas that can be found in e.g. \cite{drag-force} and \cite{herzog}.

Now consider an open string with free endpoint boundary conditions:
\begin{equation}\label{wc8}
\Pi_\mu^\sigma(0,t)=\Pi_\mu^\sigma(\pi,t)=0\, .
\end{equation}
Then, take a closed loop $\gamma$ on the string worldsheet, composed of two constant-$t$ curves at times $t_1$ and $t_2$, going from $\sigma=0$ to some chosen $\sigma=\sigma_\kappa$, connected by the two corresponding constant-$\sigma$ curves. Since the worldsheet currents are conserved, we have, again following conventions in \cite{carroll}:
\begin{eqnarray}\label{wc9}
\oint\limits_\gamma \star \Pi_\mu &=&0\\\nonumber
&=&-\int\limits_{t_1}^{t_2} dt\sqrt{-h}\,\Pi_\mu^\sigma(\sigma_\kappa,t)+\int\limits_{\sigma_\kappa}^{0} d\sigma \sqrt{-h}\,\Pi_\mu^\tau(\sigma,t_2)\\ \nonumber
&&-\int\limits_{t_2}^{t_1} dt\sqrt{-h}\,\Pi_\mu^\sigma(0,t)+\int\limits_{0}^{\sigma_\kappa} d\sigma\sqrt{-h}\,\Pi_\mu^\tau (\sigma,t_1)\, .
\end{eqnarray}
Due to the free endpoint boundary condition \eqref{wc8}, the third term on the RHS is zero, while the integrals over time, according to \eqref{wc6}, represent the spacetime momentum of the part of the string between $\sigma=0$ and $\sigma=\sigma_\kappa$ at times $t_1$ and $t_2$:
\begin{equation}\label{wc10}
p_\mu^{\sigma_\kappa}(t_2)-p_\mu^{\sigma_\kappa}(t_1)=-\int\limits_{t_1}^{t_2} dt\sqrt{-h}\,\Pi_\mu^\sigma(\sigma_\kappa,t)\, .
\end{equation}
This equation clearly shows how the momentum of some part of the string can change only if the $\Pi_\mu^\sigma$ component of the worldsheet current carries it away. The negative sign on the RHS indicates that, for a positive $\Pi_\mu^\sigma$, the momentum of that part of the string will decrease, consistent with the fact that this current component describes the flow of the momentum in the direction of increasing $\sigma$, i.e. away from the part of the string. Incidentally, if we take a string configuration which is symmetric around $\sigma_\kappa=\pi/2$, then, due to this symmetry, $\Pi_\mu^\sigma(\pi/2,t)$ must vanish. In that case $p_\mu^{\sigma_\kappa}(t)$ represents the momentum of half of the string and, from \eqref{wc10}, we see that this momentum for such a symmetric string configuration does not change with time.


\section{Energy loss}
To obtain the usual expression for the instantaneous energy loss, we can let $t_1 \to t_2$ in \eqref{wc10}:
\begin{equation}\label{en1}
\frac{dp_\mu}{dt} (\sigma,t)=-\sqrt{-h}\,\Pi_\mu^\sigma(\sigma,t)\, .
\end{equation}
This quantity gives the flow of the $\mu$ component of the momentum along the string at a position $\sigma$ at time $t$. We note again that this is the well-known expression for energy loss from \cite{drag-force} and \cite{herzog}, but the previous analysis gave us insight into its validity; namely, it pointed out that \eqref{en1} is valid only for constant-$\sigma$ curves. 

Now, let us do the following coordinate transformation on the worldsheet:
\begin{equation}\label{en2}
(\sigma,t)\to(\tilde{\sigma}(\sigma,t),t)\, ,
\end{equation}
i.e. we stay in the static gauge and only change the string parametrization using some well-defined function $\tilde{\sigma}(\sigma,t)$. We can then repeat the analysis from the previous paragraph and see that in this coordinate system, for a constant-$\tilde{\sigma}$ curve, we also have
\begin{equation}\label{en3}
\frac{d\tilde{p}_\mu}{dt} (\tilde{\sigma},t)=-\sqrt{-\tilde{h}}\tilde{\Pi}_\mu^{\sigma}(\tilde{\sigma},t)\, .
\end{equation}
We can relate these to the corresponding quantities in the $(\sigma,t)$ coordinate system by using \eqref{st7} and \eqref{st8}, which for this particular transformation are given by
\begin{eqnarray}\label{en4}
\sqrt{-\tilde{h}}&=&\frac{\sqrt{-h}}{|\tilde{\sigma}'|}\, ,\\\label{en5}
\tilde{\Pi}^{\sigma}_\mu&=&\tilde{\sigma}'\Pi^\sigma_\mu+\dot{\tilde{\sigma}}\Pi^t_\mu\, ,
\end{eqnarray}
where $\tilde{\sigma}'\equiv \partial \tilde{\sigma}/\partial\sigma$ and $\dot{\tilde{\sigma}}\equiv\partial\tilde{\sigma}/\partial t$. Plugging this in \eqref{en3} we have
\begin{equation}\label{en6}
\frac{d\tilde{p}_\mu}{dt} (\tilde{\sigma},t)={\rm sgn}(\tilde{\sigma}')\left[\frac{dp_\mu}{dt}-\sqrt{-h}\frac{\dot{\tilde{\sigma}}}{\tilde{\sigma}'}\Pi^t_\mu\right]_{(\sigma(\tilde{\sigma},t),t)}\, .
\end{equation}
If we want to evaluate the energy loss at different times, we have to make a choice of what points on the string (at different times) we are going to evaluate the currents in \eqref{en6} on. We choose that these points on the string have a constant $\tilde{\sigma}$ coordinate at all times (i.e. this is how we define the, so far, arbitrary $\tilde{\sigma}$-parametrization), while in the $\sigma$-parametrization, these points are defined by a function $\sigma_\kappa(t)$. The physical motivation behind such a choice is to say that, at some time $t$, the jet is defined as the part of the string between the endpoint $\sigma=0$ and $\sigma=\sigma_\kappa(t)$. In \cite{chesler}, for falling strings in $AdS_5$, this choice was such that the spatial distance (i.e. the $x$ coordinate in $AdS_5$, assuming that the string is moving in the $x-r$ plane) between the string endpoint and those points was of the order $\sim 1/(\pi T)$. Now, since $\tilde{\sigma}(\sigma_\kappa(t),t)$ is constant at all times, we have
\begin{equation}\label{en7}
\frac{d \tilde{\sigma}}{dt}=0=\left[\tilde{\sigma}'\frac{d\sigma_\kappa(t)}{dt}+\dot{\tilde{\sigma}}\right]_{\sigma=\sigma_\kappa(t)}\, .
\end{equation}
Plugging this in \eqref{en6} we arrive at
\begin{equation}\label{en8}
\frac{d\tilde{p}_\mu}{dt} (\tilde{\sigma},t)={\rm sgn}(\tilde{\sigma}')\left[\frac{dp_\mu}{dt}+\sqrt{-h}\,\Pi^t_\mu \frac{d\sigma_\kappa}{dt}\right]_{(\sigma_\kappa(t),t)}\, .
\end{equation}
This is the central result of this paper. This formula gives the appropriate expression for energy loss in terms of quantities expressed in any parametrization $(\sigma,t)$ in which the function $\sigma_\kappa(t)$ is known. Here we were making use of the simple expression for the energy loss in the special $\tilde{\sigma}$-parametrization (in which the coordinate of the points on which we evaluate the currents is constant), but in using this formula one does not need to know what that parametrization really is, since the RHS of \eqref{en8} is given only in terms of quantities in $(\sigma,t)$ parametrization.

Now, the argument for calling some quantity $dE/dt$ the energy loss comes from the idea that, when integrated over some period of time $\Delta t$, this integral should give the amount of energy that the jet (that is, some predefined part of the string) has lost over some period of time $\Delta t$:
\begin{equation}\label{en9}
\Delta E (\Delta t)=\int\limits_{\Delta t}dt \frac{dE}{dt}\, .
\end{equation}
By identifying $dE/dt$ with $-dp_0/dt$ in the $(\sigma=r,t)$ parametrization (essentially just the $\Pi_t^r$ component of the worldsheet current), as implied in \cite{chesler}, means that this amount of energy lost should be given by
\begin{equation}\label{en10}
\Delta E_{app} (\Delta t)=-\int\limits_{\Delta t}dt  \frac{dp_0}{dt}(r_\kappa(t),t)\, ,
\end{equation}
where the subscript {\it app} stands for ``apparent'' and where $r_\kappa(t)$ corresponds to the points at a fixed spatial distance $\sim 1/(\pi T)$ from the string endpoint at all times. However, this formula (i.e. that the energy loss is given only by the $\sigma$ component of the worldsheet current), as we showed before, is valid only if one uses a constant-$\sigma$ curve, which is not the case in this parametrization. Then, in order to be able to use that simple expression, we need to find a parametrization $\tilde{\sigma}$ in which the coordinates of the points given by $r_\kappa(t)$ are constant. In this case, the energy lost would indeed be
\begin{equation}\label{en11}
\Delta E (\Delta t)=-\int\limits_{\Delta t}dt \frac{d\tilde{p}_0}{dt}(\tilde{\sigma},t)\, .
\end{equation}
The difference between this and the apparent energy loss is then explicitly given by \eqref{en8}:
\begin{equation}\label{en12}
\Delta E (\Delta t)=\Delta E_{app} (\Delta t)-\int\limits_{\Delta t}dt\left[\sqrt{-h}\,\Pi^t_0 \frac{dr_\kappa}{dt}\right]_{(r_\kappa(t),t)}\, .
\end{equation}
In the following section we will numerically examine the effect of this correction in the case of falling strings in $AdS_5$ spacetime, dual to $\mathcal{N}=4$ SYM.


\section{Numerical evaluation}
In this section we will largely follow the procedure described in \cite{chesler}, but for consistency we will review it here. We will work in the $AdS_5$-BH geometry with the conformal boundary located at $r=0$:
\begin{equation}\label{num1}
ds^2=G_{\mu\nu}dx^\mu dx^\nu=\frac{L^2}{r^2}\left[-f(r)dt^2+dx^2+\frac{dr^2}{f(r)}\right]\, ,
\end{equation}
where $f(r)=1-r^4/r_h^4$, with $r_h$ being the radial position of the horizon of the black hole. We choose the worldsheet metric $h_{ab}$ to have the following form:
\begin{equation}\label{num2}
h_{ab}={\rm diag}(\underbrace{-s(\sigma,\tau)}_{\tau\tau},\underbrace{1/s(\sigma,\tau)}_{\sigma\sigma})\, .
\end{equation}
Here $s(\sigma,\tau)$ is the ``stretching function'', which is chosen in such a way that the numerical computation is well behaved. Choosing the worldsheet metric in this way represents merely a choice of parametrization on the worldsheet (i.e. a ``choice of gauge'') and the constraint equations \eqref{st2} are there to ensure that the embedding functions change accordingly. Explicitly, the constraint equations are
\begin{eqnarray}\label{num3}
\dot{X}\cdot X'&=&0\, ,\\ \label{num4}
\dot{X}^2+s^2(X')^2&=&0\, ,
\end{eqnarray}
where $\dot{X}^\mu\equiv \partial_\tau X^\mu$, $(X^\mu)'\equiv \partial_\sigma X^\mu$ and, in general, $A\cdot B\equiv G_{\mu\nu}A^\mu B^\nu$. We assume that the string is moving in the $x-r$ plane and choose the ``pointlike'' initial conditions, where the string is initially a point at some radial coordinate $r_c$:
\begin{equation}\label{num5}
t(\sigma,0)=0,\;\;\; x(\sigma,0)=0,\;\;\; r(\sigma,0)=r_c\, .
\end{equation}
Then $(X^\mu)'(\sigma,0)=0$, which automatically satisfies the first constraint equation \eqref{num3}, and now we have to choose an initial velocity profile (i.e. functions $\dot{X}^\mu(\sigma,0)$) such that the second constraint equation $\dot{X}^2=0$ and the free string endpoint boundary conditions \eqref{wc8} are satisfied. Following \cite{chesler}, we choose
\begin{eqnarray}\label{num6}
\dot{x}(\sigma,0)&=&Ar_c\cos(\sigma)\, ,\\ \label{num6a}
\dot{r}(\sigma,0)&=&r_c\sqrt{f(r_c)}(1-\cos(2\sigma))\, ,
\end{eqnarray}
where $A$ is a constant determining the ``amplitude'' of the velocity profile. Then $\dot{t}(\sigma,0)$ is determined by the constraint equation \eqref{num4}:
\begin{equation}\label{num7}
\dot{t}(\sigma,0)=\frac{r_c}{\sqrt{f(r_c)}}\sqrt{A^2\cos^2(\sigma)+(1-\cos(2\sigma))^2}\, .
\end{equation}
For this set of initial conditions we choose the following stretching function:
\begin{equation}\label{num8}
s(\sigma,\tau)=s(r)=\frac{1-r/r_h}{1-r_c/r_h}\left(\frac{r_c}{r}\right)^2\, .
\end{equation}
In particular, its most important feature is that it matches the singularity of the $G_{tt}$ metric component near the horizon $r_h$, so that the embedding functions can remain well behaved as parts of the string approach the horizon. 

Initial conditions \eqref{num5}, \eqref{num6} and \eqref{num6a} have been chosen as in \cite{chesler} (and in \cite{herzog}), in order to be able to exactly compare the effect of the correction on the results from that work. As discussed in \cite{chesler} and \cite{herzog}, the physical motivation behind the choice of initial conditions \eqref{num5} is that they should resemble a quark-antiquark pair produced by some local current, with the quarks having enough energy to move away in the opposite directions. The velocity profiles \eqref{num6} and \eqref{num6a} are one of the simplest profiles that satisfy the open string endpoint boundary conditions and uniformly evolve the string towards the black hole (which should resemble the process of thermalization of the interaction energy described by the body of the string) with the endpoints moving away in the opposite directions. The energy density profile in the boundary theory dual to such a string evolution will have two peaks concentrated around the endpoints, and the smooth U-shaped profile between the peaks will slowly decrease in magnitude.

\begin{figure}[h!]
	  \centering
    \includegraphics[width=0.47\textwidth]{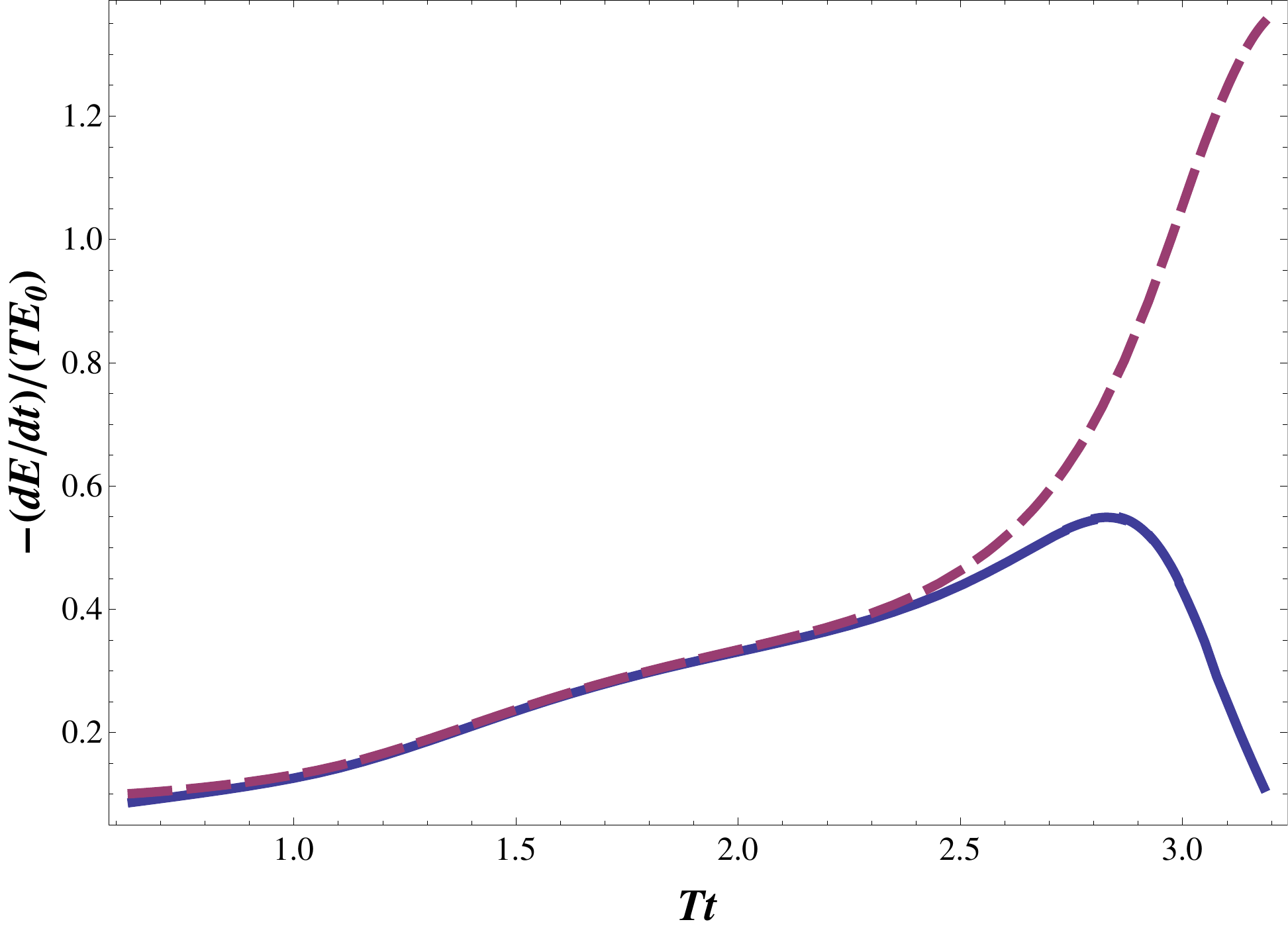}
    \caption{\label{fig}Comparison of the (normalized) instantaneous energy loss as a function of time with and without the correction in \eqref{en8}. The dashed red curve shows the apparent energy loss $(dE/dt)_{app}$ in the radial $\sigma=r$ parametrization (Eq. \eqref{extra2}), while the solid blue curve is the actual energy loss $dE/dt$, as given by \eqref{extra1}. The energy loss was evaluated at points at a fixed spatial distance from the string endpoint, chosen in such a way that the correction in \eqref{en8} appears clearly. The normalization constant $E_0$ is the energy of half of the string and $T=1/(\pi r_h)$ is the temperature. The numerical parameters used are $A=50$ and $r_c=0.1r_h$.}
\end{figure}

With this choice of initial and boundary conditions, we can solve the equations of motion \eqref{st3} numerically, obtain the embedding functions $X^\mu(\sigma,\tau)$ and then evaluate the energy loss in the radial $\sigma=r$ parametrization with and without the correction in \eqref{en8}. To obtain the actual energy loss, we simply use the formula \eqref{en8} with the worldsheet fluxes expressed in the static gauge $(\sigma,t)$ using \eqref{st7} and \eqref{st8}:
\begin{equation}\label{extra1}
\frac{dE}{dt}=\frac{L^2}{2\pi\alpha'}\left[\frac{f}{r^2}\frac{1}{|\dot{t}|}\left(st'-\frac{d\sigma_\kappa}{dt}\left(s\left(t'\right)^2-\frac{\dot{t}^2}{s}\right)\right)\right]_{(\sigma_\kappa(t),t)}\, .
\end{equation}
The apparent energy loss is given by \eqref{en1} in the $(r,t)$ parametrization, so we need to use formulas \eqref{st7} and \eqref{st8} again, giving
\begin{equation}\label{extra2}
\left(\frac{dE}{dt}\right)_{app}=\frac{L^2}{2\pi\alpha'}\left[\frac{f}{r^2}\frac{sr't'-\frac{1}{s}\dot{r}\dot{t}}{\left|r'\dot{t}-\dot{r}t'\right|}\right]_{(\sigma_\kappa(t),t)}\, .
\end{equation}
The results are shown in Fig. \ref{fig}. One can clearly see that the correction, derived in \eqref{en8}, becomes especially important at late times, when $dr_\kappa/dt$ grows, as the relevant parts of the string start falling towards the black hole faster and faster. 

We should note that the energy of the part of the string between $\sigma=0$ and $\sigma=\sigma_\kappa(t_{th})$ at the thermalization time $t_{th}$ (when the string endpoint stops moving in the $x$-direction), is generally nonzero. This means that the area under the solid blue curve in Fig. \ref{fig} is always less than 1, and represents the relative amount of energy lost from the part of the string defined by $\sigma_\kappa(t)$, as evident from \eqref{wc10}. On the other hand, the area under the dashed red curve is not known {\it a priori}, and could be $<1$ or $>1$, depending on the magnitude of the correction in \eqref{en8}. Specifically, if we decrease the spatial distance from the endpoint at which we evaluate the energy loss (keeping the same initial conditions), the area under the red curve increases and eventually becomes $>1$ (in the figure it is already slightly higher than 1). 

Also note that the ``jet definition'' we are using here is taken from \cite{chesler}, where the jet is defined as the part of the string within a certain $\Delta x$ distance from the endpoint. As discussed in \cite{chesler}, the physical motivation behind this is that the baryon density in the boundary theory should be well localized on scales of order $\Delta x\sim1/{\pi T}$. There are also other physically well-motivated jet definitions \cite{will-razieh}, exploration of which is left for future work.


\section{Discussion}
We have derived, by analyzing transformations of spacetime momentum fluxes on the classical string worldsheet, a general expression \eqref{en8} for calculating the instantaneous energy loss for time-dependent string configurations valid in any choice of worldsheet parametrization. This formula shows that the energy loss in time-dependent string configurations receives a correction to the simple $\Pi_\mu^\sigma$ expression. This correction comes from the fact that the points on the string at which we want to evaluate the energy loss at different times do not necessarily have constant coordinates in the chosen worldsheet parametrization. The importance of the correction depends on how fast the coordinates of these points change in time in that parametrization, i.e. on the magnitude of the  $d\sigma_\kappa(t)/dt$ function. In the example of falling strings, we have seen that this correction becomes especially important at late times and substantially decreases the magnitude of the Bragg-like peak (Fig. \ref{fig}). 

One should point out that this correction does not affect the results of \cite{chesler} for the maximum stopping distance $(\Delta x)_{max}\sim E^{1/3}$, since this expression was derived from purely kinematical considerations, by analyzing the equations of motion and relating the total energy of the string to the approximate endpoint motion described by the null geodesics. In other words, the worldsheet currents (actually, their identification with the energy loss) were not used in that derivation.

We should also point out that this correction does not affect the well-established drag force results of \cite{drag-force} and \cite{herzog}, since the trailing string is a stationary string configuration where $d\sigma_\kappa/dt=0$.

One is tempted to speculate about the implications of the results shown in Fig. \ref{fig} to the jet quenching phenomena. Our preliminary numerical studies suggest that, although the early time behavior of the energy loss is susceptible to the initial conditions (as noted in \cite{chesler}), the linearity of it seems to be a remarkably robust feature. Of course, a more thorough numerical analysis is needed to confirm such a claim (and is left for future work), but if this indeed remains to be true and $dE/dt$ scales like $\sim t^1$, then taking into account that $(\Delta x)_{max}\sim E^{1/3}$, it can be shown that this is similar to the typical qualitative behavior of energy loss of light quarks in pQCD in the strong LPM regime \cite{miklos-abc}. This suggests a tempting idea that the phenomenon of light quark jet quenching may have a roughly universal qualitative character, regardless of whether we are dealing with a strongly or a weakly coupled medium. 

However, as emphasized before, a more thorough quantitative analysis and estimate of the relative magnitude of energy loss and stopping distances are necessary. It would be also interesting to more thoroughly examine the effects of varying the string initial conditions, as well as choosing a different $\sigma_\kappa(t)$ function (i.e. a different jet definition), on the shape of the instantaneous energy loss. By choosing some set of these, one can then compute the $R_{AA}$ in a dynamical, expanding medium with a realistic set of nuclear initial conditions, and, finally, inspect the robustness of that result by varying the string initial conditions and the $\sigma_\kappa(t)$ choice and seeing how much $R_{AA}$ would be affected.

\begin{acknowledgments}
We especially thank J. Noronha and M. Gyulassy for valuable input and help. We also thank M. Mia, S. Endlich, A. Buzzatti, C. Herzog and S. Gubser for helpful discussions. We acknowledge support by U.S. DOE Nuclear Science Grant No. DE-FG02-93ER40764.
\end{acknowledgments}

\end{document}